\documentclass[prl,showpacs,preprint,preprintnumbers,amsmath,amssymb,superscriptaddress]{revtex4}

\usepackage{amsmath,graphicx,latexsym,times,color}
\usepackage{setspace}

\begin{document}

\title{Deriving molecular bonding from macromolecular self-assembly}

\author{Fabien Silly}
\affiliation{Department of Materials, University of Oxford, Parks
Road, Oxford OX1 3PH, UK.}
\affiliation{Zernike Institute for Advanced Materials, University of Groningen, Nijenborgh 4, NL-9747 AG, Groningen, The Netherlands}
\author{Ulrich K. Weber}
\affiliation{Department of Materials, University of Oxford, Parks
Road, Oxford OX1 3PH, UK.}
\author{Adam Q. Shaw}
\affiliation{Department of Materials, University of Oxford, Parks
Road, Oxford OX1 3PH, UK.}
\author{Victor M. Burlakov}
\affiliation{Department of Materials, University of Oxford, Parks
Road, Oxford OX1 3PH, UK.}
\affiliation{Institute for Spectroscopy Russian Academy of Sciences, Troitsk 142190, Russia.}
\author{Martin R. Castell}
\affiliation{Department of Materials, University of Oxford, Parks
Road, Oxford OX1 3PH, UK.}
\author{G. A. D. Briggs}
\affiliation{Department of Materials, University of Oxford, Parks
Road, Oxford OX1 3PH, UK.}
\author{David G. Pettifor}
\affiliation{Department of Materials, University of Oxford, Parks
Road, Oxford OX1 3PH, UK.}\

\date{\today}

\begin{spacing}{2}  

\begin{abstract}

Macromolecules can form regular structures on inert surfaces. We have developed
a combined empirical and modeling approach to derive the bonding. From
experimental scanning tunneling microscopy (STM) images of structures formed on Au(111) by melamine, by PTCDA,
and by a 2:3 mixture of the two, we determine the molecular bonding
morphologies. Within these bonding morphologies and recognizing the distinction
between cohesive and adhesive molecular interactions we simultaneously
simulated different molecular structures using a lattice Monte Carlo method.
Within these bonding morphologies there is a distinction between cohesive and adhesive molecular interactions. We have simulated different molecular structures using a lattice Monte Carlo method.

\end{abstract}



\newcommand{\Ea}{\ensuremath{E_a}}
\newcommand{\Eb}{\ensuremath{E_b}}
\newcommand{\Ec}{\ensuremath{E_c}}
\newcommand{\Ed}{\ensuremath{E_d}}
\newcommand{\Ee}{\ensuremath{E_e}}

\newcommand{\eV}{\ensuremath{\,eV}}

\maketitle

Self-assembly of molecules on atomically well-defined surfaces offers a
bottom-up approach for generating two-dimensional nanostructures
\cite{Barth05,Lopinski00,Zhang07,Chen08,Silly07,Dusastre00,Tanev95,Silly08a,Theobald03,Bonifazi07,Winfree98,Barth07,Otero05,Silly08b}.
To develop structures for specific purposes requires a precise knowledge of the
molecular bonding, including molecular binding rules and the corresponding
binding energies \cite{Franke08,Yokoyama01}. These cannot always be calculated ab initio
because of the complex character of molecular bonding and the difficulty of
including dispersive (van der Waals) interactions \cite{Grimme04}.

Hydrogen bond-forming molecules are particularly suitable for generating
self-assembled structures due to the high selectivity and directionality of
hydrogen bonds \cite{Conn97,Archer01,Prins01}, and relatively low energetics
which enable equilibrium molecular configurations to be achieved at relatively
low processing temperatures  \cite{Theobald03,Swabrick06}. Using mixtures of
different hydrogen bond-forming molecules, with diverse binding rules,
\cite{Kelly05,Kelly05b} allows the formation of a wider variety of molecular
structures by changing molecular composition, and presents a promising approach
to generating molecular scaffolds \cite{Moriuchi04}. The relation between the
properties of individual molecules and the characteristics of their
self-assembly on surfaces remains to be understood. Our study provides a step
towards establishing such a relation by solving the inverse problem -
extracting characteristics of molecular interactions by analysing
self-assembled structures.

In this paper we use experimental observations of a self-assembled molecular structure to determine
plausible binding rules, and then perform kinetic Monte Carlo simulations to
estimate the binding energies. We analyse simultaneously the structural
stability of the stoichiometric molecular mixture and the individual
constituent molecular components. We apply this methodology to PTCDA and
melamine molecules on Au(111), where the substrate has little effect on
intermolecular interactions.

Our procedure for extracting the molecular binding rules and the corresponding
interaction energies involves two major steps. In the first step we analyse
experimentally observed molecular structures in order to extract the
characteristics of the molecular arrangement for all molecular compositions of
interest. Thereby we identify the ways molecules bind to each other, i.e. the
molecular binding rules. In the second step we construct the model system using
the binding rules and arranging the molecules on a grid, suitable for
on-lattice kinetic Monte Carlo simulations of molecular structures. Thereby we
extract the molecular binding energies by analysing simultaneously the
stability of both the single-component molecular structures and of that formed
from a binary mixture.

We used Au(111) film grown on mica
substrates. The samples were introduced into the ultrahigh vacuum (UHV) chamber
of a STM (JEOL JSTM4500S) operating at a pressure of 10$^{-8}$ Pa. The Au(111)
surfaces were sputtered with argon ions and annealed in UHV at temperatures
between 600 and 800$\,^{\circ}\mathrm{C}$ typically for 30 min. PTCDA molecules
were sublimated at 275$\,^{\circ}\mathrm{C}$ and melamine at
100$\,^{\circ}\mathrm{C}$. Electrochemically etched tungsten tips were used to
obtain constant current (I$_t$) images at room temperature with the bias
voltage (V$_s$) applied to the sample. The structures of molecular mixture was
obtained after deposition of PTCDA on Au(111), followed by a deposition of
melamine with a 2:1 ratio and a post annealing at 90$\,^{\circ}\mathrm{C}$ for
10 hours.


\begin{figure}
\includegraphics[scale=0.45]{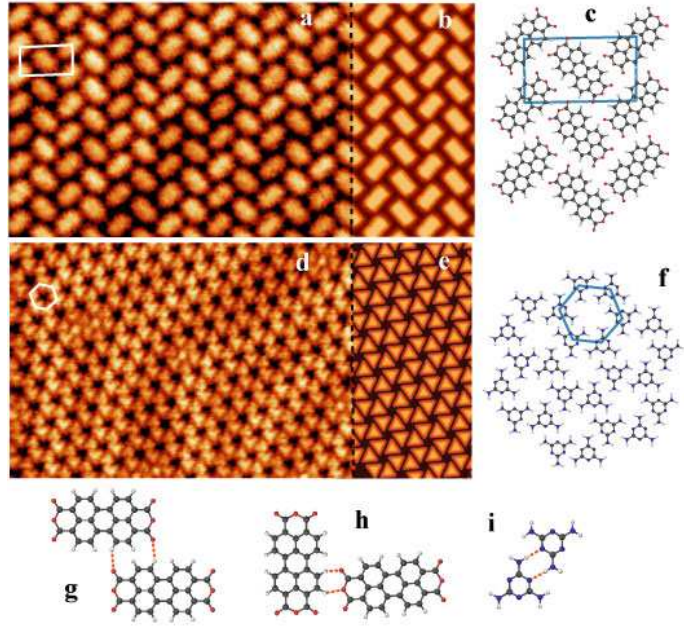}
 \caption{\label{fig:fig1}
 (a-c) PTCDA domain: a) STM-image on Au(111) surface
(14$\times$10nm$^2$; V$_s$ = +1.5 V, I$_t$ = 0.4 nA) , b) simulation result
displayed on a sheared hexagonal lattice, c) molecular ordering. (d-f)
Melamine domain: d) STM-image (14$\times$10nm$^2$; V$_s$ = -1.0 V, I$_t$ = 0.5
nA), e) simulation shown on a simple hexagonal lattice, f) molecular ordering.
The unit cell is outlined in blue. (g,h,i) Plausible molecular bonds
occurring in structures c and f. In the molecule 3D representation, gray balls
are carbon atoms, red balls are oxygen atoms, white balls are hydrogen atoms
and blue balls are nitrogen atoms.}
\end{figure}

Molecular binding rules between identical molecules are obtained by comparing
the molecular arrangements in the single-component molecular structures. The
experimental images of the molecular structures of
3,4,9,10-perylene-tetracarboxylic-dianhydride (PTCDA) and
1,3,5-triazine-2,4,6-triamine (melamine) self-assembled on Au(111) are shown in
Fig. \ref{fig:fig1}. Fig.\ref{fig:fig1}a shows the compact domains of PTCDA, which exhibit a uniform
structure with a herringbone-like pattern. The 12.0 $\times$ 20.0 \AA$^2$ unit
cell of the structure is rectangular and contains two molecules with their main
axes oriented at an angle of 86$\,^{\circ}$ with respect to each other \cite{Schmitz97}.
Molecular bonding in the PTCDA structure can be characterized by two
independent parameters  E$_{pp}^{(1)}$ and E$_{pp}^{(2)}$  as illustrated in
Fig. \ref{fig:fig1}g and 1h, respectively. Melamine molecules on Au(111), according to Fig.
1d, form domains of chiral structure and hexagonal symmetry with the lattice
parameter of 9.8 \AA. This arrangement is stabilized by a double hydrogen bond
\cite{Xu07}, as illustrated in Fig. \ref{fig:fig1} i. The corresponding energy E$_{mm}$ for
the melamine-melamine bonding is found to be  E$_{mm}$ = 0.45 eV
\cite{Weber07}.

\begin{figure}
\includegraphics[scale=0.46]{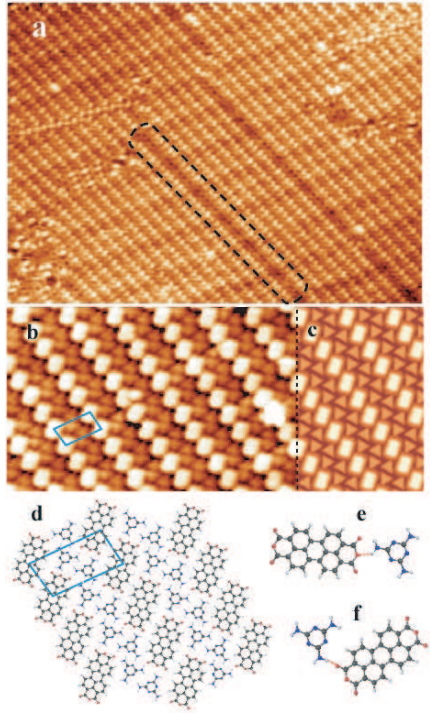}
 \caption{\label{fig:fig2}
 Mixed PTCDA and melamine domain. (a) STM-image on Au(111) surface
(80$\times$60nm$^2$; V$_s$ = -1.2 V, I$_t$ = 0.2 nA) , (b) Close up
(14$\times$8 nm$^2$; V$_s$ = -1.5 V, I$_t$ = 0.1 nA) , (c) simulation result
shown on a sheared hexagonal lattice, (d) model of the molecular ordering,
(e-f) plausible molecular bonds occurring in d.}
\end{figure}

Details of the melamine-PTCDA molecular bonding are obtained by analysing the
structure of the melamine-PTCDA mixture with the composition 2:1. This large
scale structure is formed of ordered molecular stripes, as shown in Fig. \ref{fig:fig2}a.
Each stripe is composed of a single PTCDA molecular row and a double row of
melamine molecules, Fig. \ref{fig:fig2}b. The PTCDA molecular axis is rotated by
50$\,^{\circ}$ with respect to the stripe line thereby making this structure
chiral. The unit cell outlined in Fig. \ref{fig:fig2}b in blue has a parallelogram shape,
with an angle of 85$\,^{\circ}$ with 10.0 \AA~ (the periodicity along the PTCDA rows) and
19.9 \AA~ parameters (PTDCA-PTCDA separation across 2 melamine rows). Fig. \ref{fig:fig2}d
shows the molecular arrangement as observed in Fig. \ref{fig:fig2}b. Molecular interactions
in the PTCDA-melamine structure can in general be characterized by two
parameters E$_{mp}^{(1)}$  and E$_{mp}^{(2)}$  corresponding to physically
reasonable hydrogen bonding as illustrated in Fig. \ref{fig:fig2}e,f.

The molecular binding energies  E$_{pp}^{(1)}$,  E$_{pp}^{(2)}$, E$_{mp}^{(1)}$
and  E$_{mp}^{(2)}$ can now be estimated by studying the stability of the
self-assembled single-components and the mixed molecular structure using the
kinetic Monte Carlo methodology \cite{Weber07} with an underlying hexagonal
grid defining the topology of molecular movements and interactions and an
(N,V,T) ensemble.

\begin{figure}
\includegraphics[scale=0.5]{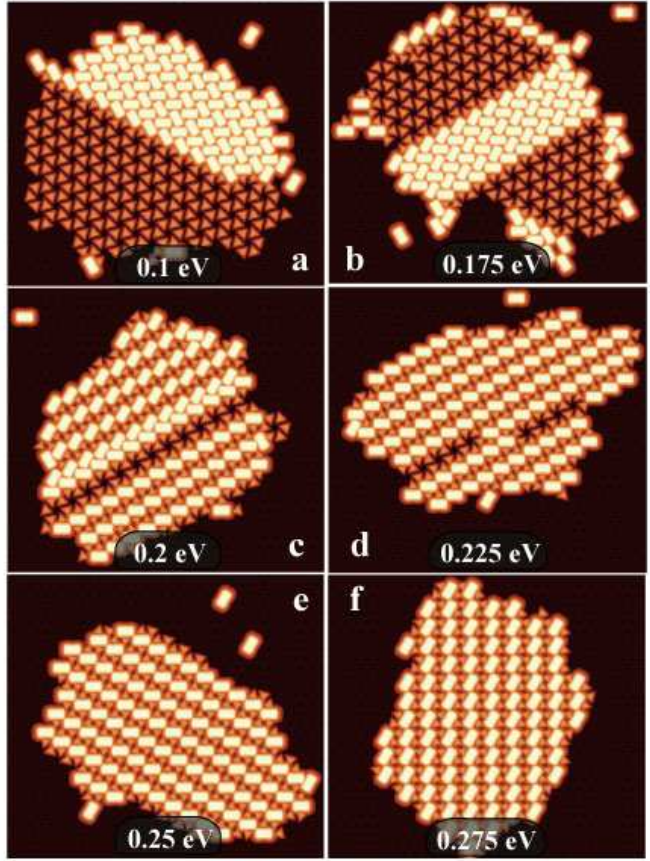}
\caption{\label{fig:bonds} Simulated PTCDA-melamine ordering depneding on
PTCDA-melamine bond energies $E_{mp}^{(1)}=E_{mp}^{(2)}=E_{ad}$, varying from
$0.1 eV$ to $0.275 eV$. (a,b) Small bond energies between PTCDA and melamine
with $E_{ad}\leq0.175$ lead to a phase segregation. (c,d) In a narrow energy
range of $0.2\leq E_{ad}\leq 0.225$ melamine double rows can be observed as a
typical defect. (e-f) Energies $E_{ad}\geq0.25$ allow the formation of a
defect-free PTCDA-melamine domain.}
\end{figure}

\begin{figure}
\includegraphics[scale=0.6]{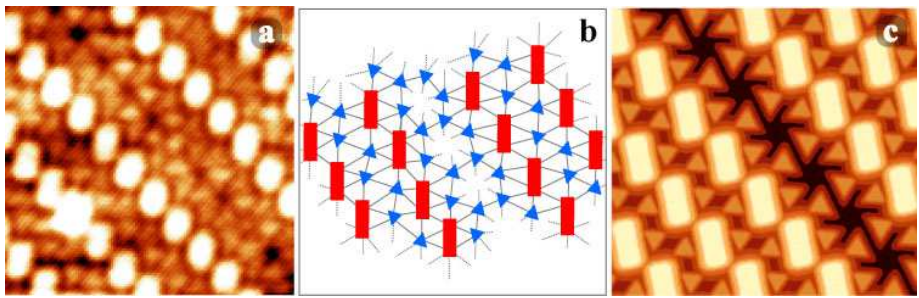}
 \caption{\label{fig:vergleich}
 The double melamine row in an PTCDA-melamine
domain.
(a) close-up of an STM image (8$\times$8nm$^2$; V$_s$ = -1.2 V, I$_t$ =
0.2 nA).
(b) Scheme showing the common topology of both experiment and
simulation. Red rectangles are PTCDA molecules and blue triangles are
melamine molecules.
(c) Simulated image using a simple hexagonal geometry.}
\end{figure}

The on-lattice kinetic Monte Carlo model is an entirely topological model. It
exclusively considers binding energies between molecules that are nearest
neighbours in the underlying lattice, and does not contain information about
the absolute position and orientation of molecules. Therefore the point of
comparison between experimental and simulated structures is the topological
correspondence (Fig. \ref{fig:vergleich}). The freedom to apply different geometries to a
simulation while keeping the topology unchanged can be used for visualisation
of highly ordered homogeneous molecular arrangements (Fig.\ref{fig:fig1}, Fig.\ref{fig:fig2}), where a
sheared hexagonal geometry achieves a good correspondence. In other cases where
the geometry is more complex due to defects or irregularities (Fig. \ref{fig:bonds}, Fig. \ref{fig:vergleich}), the simulation is presented with a simple hexagonal geometry. In each case, what
matters is the correspondence in topology between simulation and experiment.

In our simulations we use a binary mixture of anisotropic molecules of type 1
(trigonal vertices) and type 2 (linear rods) occupying one and two sites,
respectively, on a 30$\times$30 two-dimensional hexagonal lattice, initialized in a
random configuration. Rod-like molecules exist in three different orientations
along the symmetry axes of the hexagonal grid, while vertex-like molecules
exist in 2 different orientational configurations, due to their two-fold and
three-fold molecular symmetry, respectively. Any pair of molecules that exists
in one of the nearest neighbour configurations, shown in Fig. \ref{fig:fig1}g,h,i and
Fig. \ref{fig:fig2} e,f, establishes a hydrogen bond with the energies  E$_{mm}$,
E$_{pp}^{(1,2)}$ , E$_{mp}^{(1,2)}$, respectively. The number of molecules (200
vertices, 100 rods) is chosen to be stoichiometric, complying with the ratio of
the experimentally observed unit cell shown in Fig. \ref{fig:fig2}. The simulation
temperature of k$_BT$=0.08 eV is higher than in experiment to enable short
equilibration times, but low enough to avoid thermally generated defects.

In our simulations we neglect the influence of the substrate on the molecular
binding energies, which is a reasonable approximation for the PTCDA and
melamine molecules on Au(111) \cite{Henze07}. The structure of the molecular mixture can
be unstable against phase separation into single-molecular domains. This
instability is controlled by the strength of the cohesive interactions
associated with E$_{pp}^{(1)}$ and E$_{pp}^{(2)}$  relative to the adhesive
interactions associated with  E$_{mp}^{(1)}$ and E$_{mp}^{(2)}$. This trade-off can be simplified by assuming that
E$_{pp}^{(1)}$$\approx$E$_{pp}^{(2)}$=E$_{coh}$  and
E$_{mp}^{(1)}$$\approx$E$_{mp}^{(2)}$=E$_{adh}$ reducing the number of fit
parameters down to two.

In Fig. \ref{fig:fig1} and Fig. \ref{fig:fig2} we show the predicted structures with high symmetry molecular
ordering in order to compare with the illustrated experimental images. The pure
PTCDA domain in Fig. \ref{fig:fig1}b and the pure melamine domain in Fig. \ref{fig:fig1}e are both the
result of the binding rules and the high enough binding energies to ensure
thermal stability of the structures. The stability of the PTCDA-melamine
structure requires a certain range of adhesive interaction energies for given
values of the cohesive interaction energies, as illustrated in Fig. \ref{fig:bonds}.

In Fig. \ref{fig:bonds}a-f we show how the structure undergoes a change, caused by varying
the PTCDA-melamine interaction strengths
E$_{mp}^{(1)}$=E$_{mp}^{(2)}$=E$_{ad}$. We observe that values of E$_{ad}$ =
0.175 eV, E$_{coh}$ = E$_{pp}^{(1)}$=E$_{pp}^{(2)}$= 0.15 eV lead to the phase
separation due to the weak PTCDA-melamine interactions. Whereas E$_{ad}$ =
0.2 eV creates structures with long range ordering and a sporadic but
reproducible occurrence of double melamine row defects. Fig. \ref{fig:vergleich} highlights that
these double row defects in the PTCDA-melamine domains are observed in STM
(Fig. \ref{fig:fig2}a and Fig. \ref{fig:vergleich}a) and are also predicted by our simulation (Fig. \ref{fig:vergleich}c). It can
be seen from Fig. \ref{fig:vergleich}b that both experiment (Fig. \ref{fig:vergleich}a) and simulation (Fig. \ref{fig:vergleich}c)
correspond topologically. Reproducing the double-row defect in our simulations
allows the range for the effective energy parameters to be narrowed as given in
Table 1.

\begin{table*}
\begin{centering}
\begin{tabular}{@{\quad}l@{\qquad}c@{\qquad}cc@{\qquad}c@{\quad}}
\hline
 \rule[-1.5ex]{0mm}{4ex}    Type & Bond type    & Energy    & Arrangement & Energy range \tabularnewline
\hline\hline $mm$    & 2 $\times N-H\cdots O$     & $E_{mm} = 0.45 eV$\quad
\cite{Weber07} & \parbox{2cm}{\includegraphics[width=1.8cm]{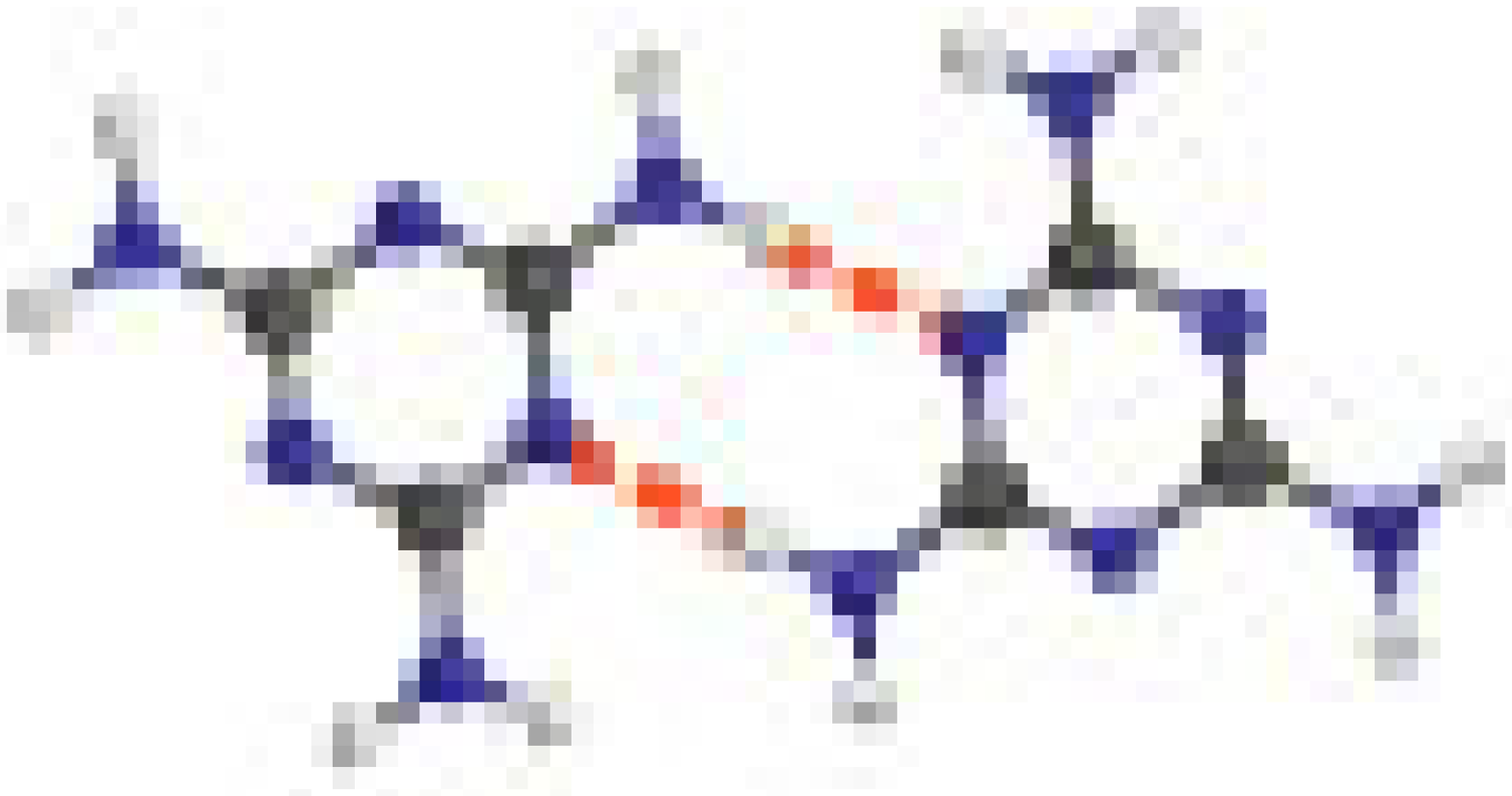}} &
{\scriptsize reference value} \tabularnewline \hline
 $mp^{(1)}$ &   $N-H\cdots O$ & $E_{mp}  =E_{adh}= 0.225 eV$ & \parbox{3cm}{\includegraphics[width=2.4cm]{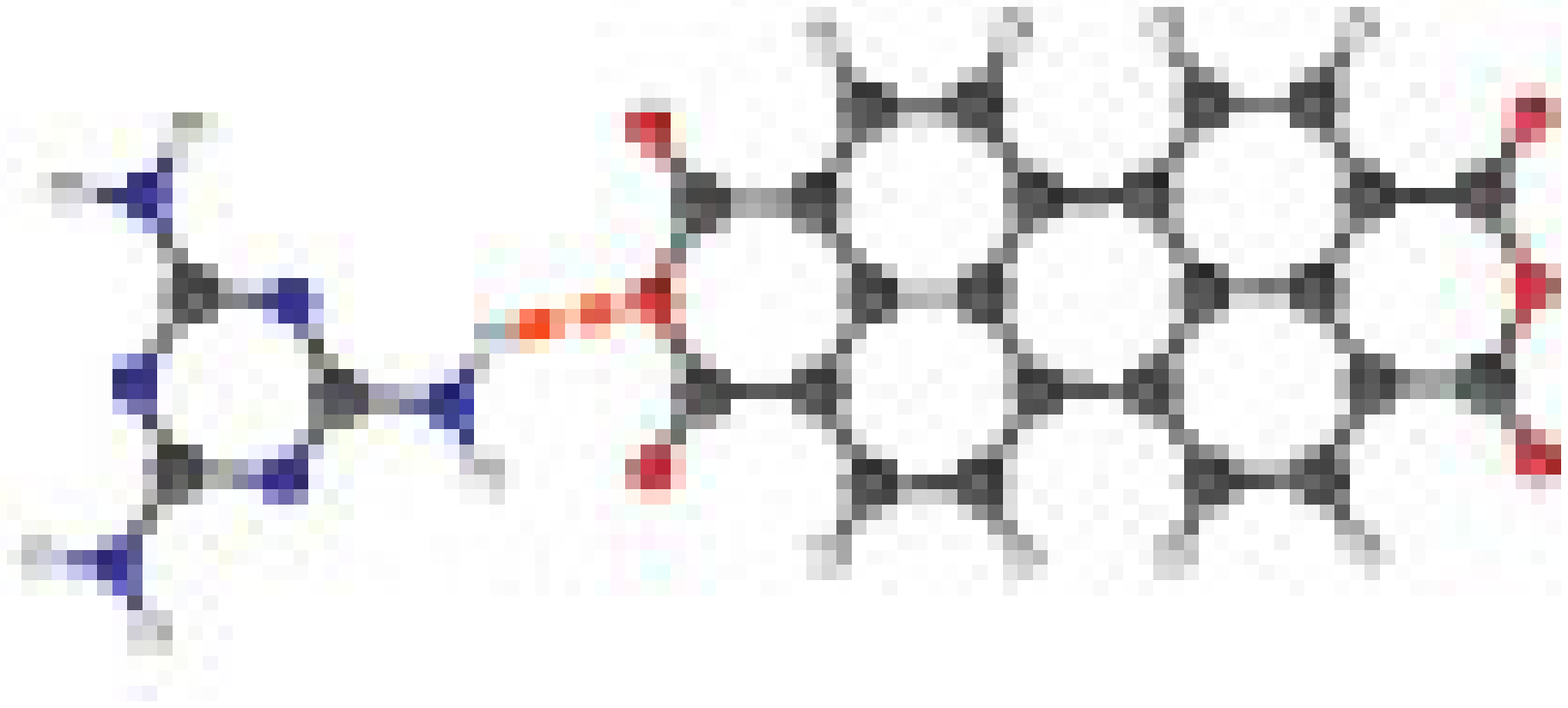}}&  $0.2 eV <  E_{adh}  < 0.25 eV$ \tabularnewline
\hline
 $mp^{(2)}$ & $N-H\cdots O$ & $E_{mp} = E_{adh}= 0.225 eV$   &\parbox{2cm}{\includegraphics[width=1.8cm]{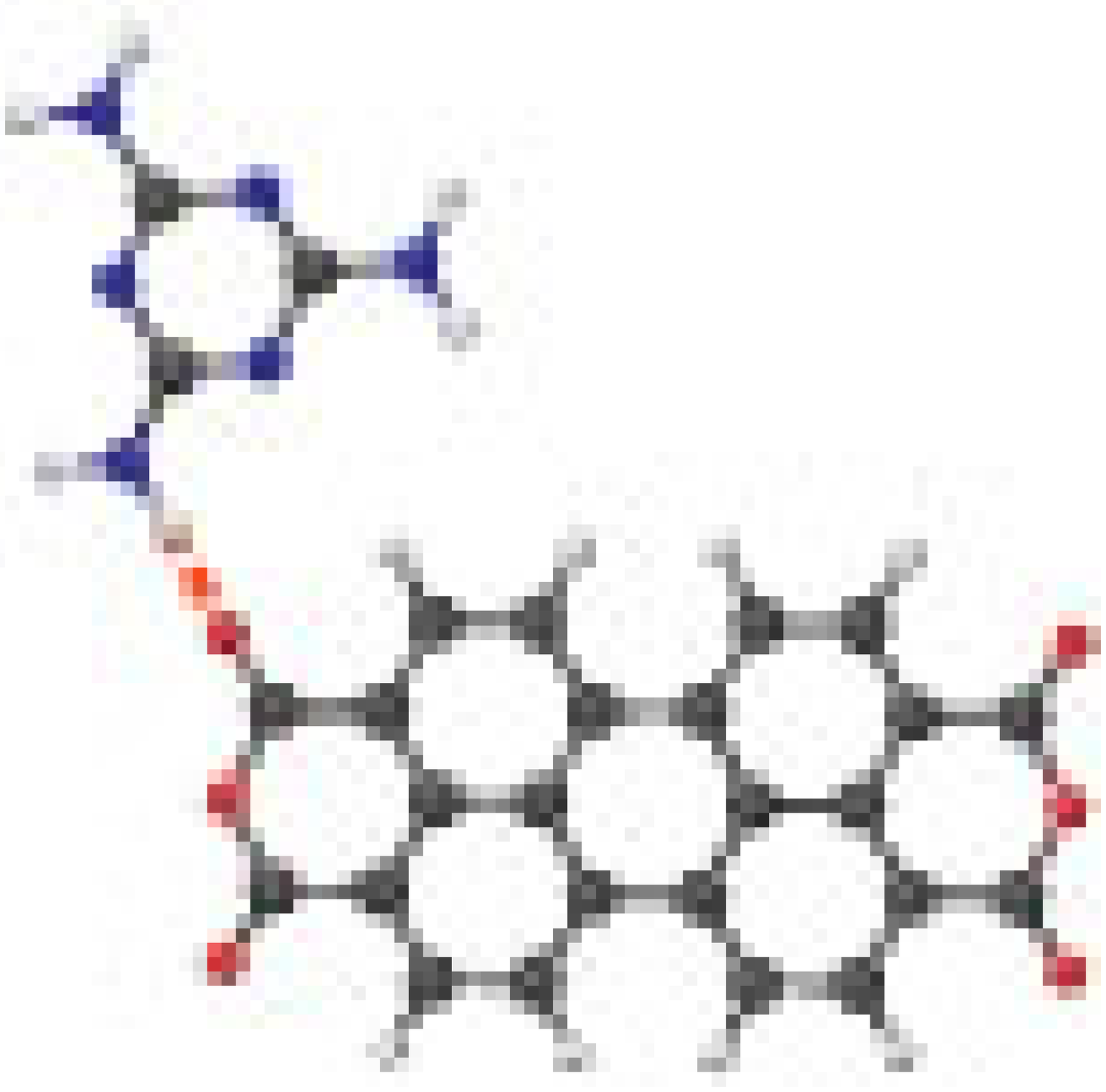}} &  $0.2 eV < E_{adh} < 0.25 eV$ \tabularnewline
\hline
$pp^{(1)}$ &    $2 \times  C-H\cdots O$ &    $E_{pp} = E_{coh}= 0.15 eV$  & \parbox{2cm}{\includegraphics[width=1.8cm]{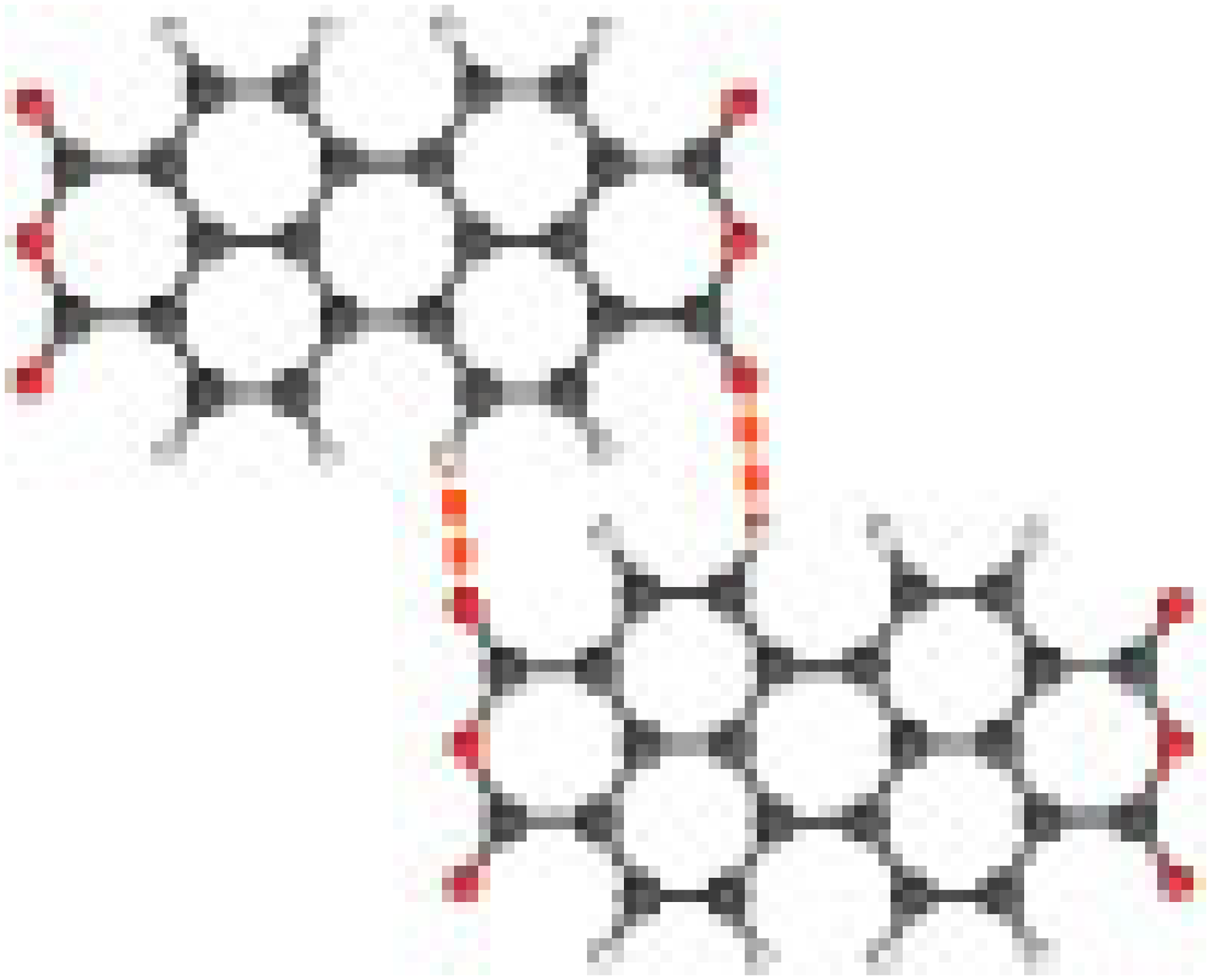}}& $0.1 eV <  E_{coh} < 0.3 eV$ \tabularnewline
\hline
$pp^{(2)}$  & $2 \times  C-H\cdots O$   &  $E_{pp} = E_{coh} = 0.15 eV$     & \parbox{2cm}{\includegraphics[width=1.8cm]{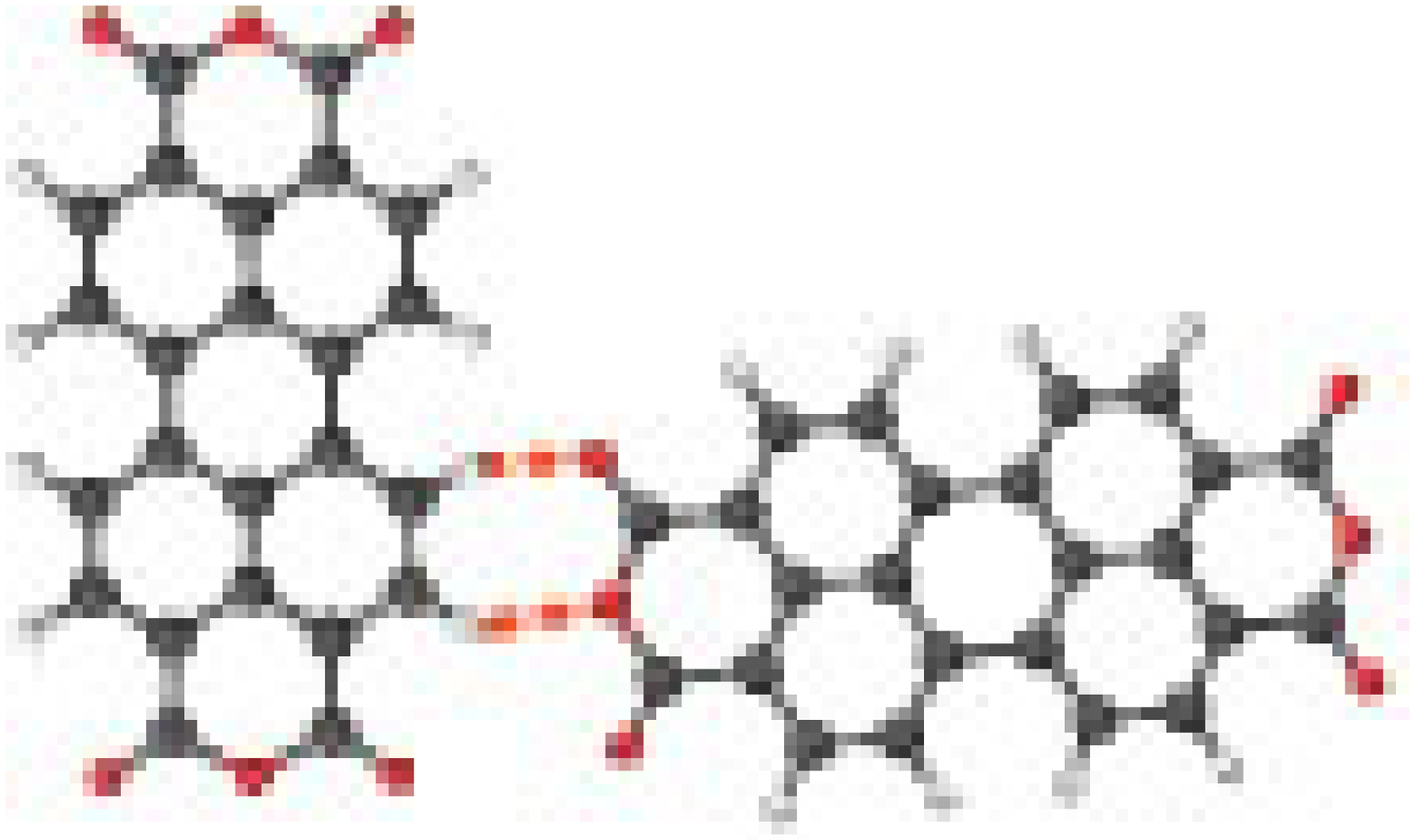}}& $0.1 eV <  E_{coh} < 0.3 eV$ \tabularnewline
\hline

\end{tabular}
\par\end{centering}

\caption{Summary of binding rules and energies in the PTCDA-melamine system.
Abbreviation: mm=(melamine-melamine), pp=(PTCDA-PTCDA) and mp=(melamine-PTCDA).
\label{tab:Lattice1}}

\end{table*}

The interaction energy values we found allow all three experimentally observed
structures shown in Fig. \ref{fig:fig1} and Fig. \ref{fig:fig2} to be simulated simultaneously, as illustrated in Fig. \ref{fig:fig1}b,e, and Fig. \ref{fig:fig2}c. The obtained energies  E$_{ad}$ and E$_{coh}$ comprise the contributions from both the hydrogen bonds and van der Waals interactions. The
latter energies are usually in the range 0.04-0.1 eV \cite{Grimme04}, which is
around the lower limit of the variation range identified for the parameters
E$_{ad}$ and E$_{coh}$. Hence the major contribution to the obtained values of
E$_{ad}$ and E$_{coh}$ can be associated with the hydrogen bond energies.


By combining STM images of molecular ordering in a PTCDA-melamine system with
on-lattice Monte Carlo simulations of the structural stability we have
determined the molecular binding energies. We have decomposed the molecular
interactions into adhesive and cohesive parts, which reduces the number of free
parameters and hence the uncertainty in the energy values. The results provide
information which can be used as a starting point for more detailed studies of
molecular bonding using more sophisticated calculation techniques. This
procedure of extracting the molecular binding rules and estimating binding
energies is applicable to a wide range of multicomponent systems.

\textbf{Acknowledgment.}  The authors thank the EPSRC (EP/D048761/1 and
GR/S15808/01) for funding and Chris Spencer (JEOL UK) for valuable technical support.

\narrowtext

\end{spacing}

\end{document}